\documentstyle[12pt]{article}
\date{ }

\title{Field-enlarging transformations and chiral theories}

\author{J. S\l adkowski $^{\dag}$  \\
Fakult\" at f\" ur   Physik Universit\"at Bielefeld,\\
D-4800 Bielefeld 1, Universit\"atsstrasse 25, Germany}
\begin{document}
\baselineskip9mm
\maketitle
\begin{abstract}
\baselineskip9mm
   A field-enlarging transformation in the chiral electrodynamics is
performed. This introduces an additional gauge symmetry to the model
that is unitary and anomaly-free and allows for comparison of different
models discussed in the literature.  The  problem  of  superfluous
degrees of
freedom and their influence on quantization is discussed. Several
"mysteries" are explained from this point of view.

\end{abstract}
\vspace{40mm}

$^{\dag}$ A. von Humboldt Fellow; permanent address: Dept.
of Field Theory and
Particle Physics, University of Silesia, Pl 40007 Katowice, Poland.

\newpage
\ \ \ Consistent quantization of an anomalous chiral gauge theory have,
for a long time, been problematic. In several simple cases, a physically
consistent and unitary models can be obtained $^{1-7}$. But it still
remains one of the most important open questions in field theory $^{5, 8, 9}$.
To solve the problem one usually adds in a more
or less sophisticated way additional terms to the Lagrangian $^{1-7,10-12}$.
Another way is to introduce a non-local gauge-fixing or
interaction term $^{13-15}$. The resulting theory
is then invariant with respect to a restricted gauge symmetry that is not
anomalous. Here we would like to apply a field-enlarging transformation
to analyse the problem $^{3,16-18}$. This transformation introduces additional
scalar degrees of freedom to the system and restores gauge symmetry,
although not always the one one started with.
It is then possible to show explicitely the relations among various
proposals and how the mechanism works. The conventional common
part of Lagrangian for the discussed in the literature models
is (chiral electrodynamics):

$$ {\it L} = - \frac{1}{4} F^{\mu \nu} F_{\mu \nu} + {\bar \psi}\left[
i \partial ^{\mu} \gamma _{\mu} - \frac{e}{2} \left( 1 + \gamma ^{5} \right)
A{_\mu} \gamma ^{\mu} \right] \psi \ . \eqno(1)$$
This Lagrangian is invariant with respect to

$$\delta A_{\mu}=\partial _{\mu} \alpha  \eqno(2a)$$

$$\delta \psi = -i\alpha \frac{e}{2} \left( 1-\gamma ^{5}\right)
\psi \eqno(2b)$$

$$\delta {\bar \psi} = i \frac{\alpha e}{2}{\bar \psi} \left(1+ \gamma ^{5}
\right) \ , \eqno(2c) $$
where $\alpha$ is an arbitrary real function. Unfortunately,
this gauge invariance is spoiled at the quantum level $^{19}$.
Let us perform the following field-enlarging
transformation $^{3, 16-18}$

$$ A_{\mu} \rightarrow A_{\mu} - \partial _{\mu} \phi \equiv
g_{\mu} \left( A, \phi\right) \eqno(3)  $$
in the Lagrangian (1). The transformed Lagrangian has the form

$$ {\it L} = - \frac{1}{4} F^{\mu \nu} F_{\mu \nu} + {\bar \psi}\left[
i \partial ^{\mu} \gamma _{\mu} - \frac{e}{2} \left( 1 + \gamma ^{5} \right)
\left( A_{\mu} \gamma ^{\mu} - \gamma ^{\mu}
\partial _{\mu} \phi \right) \right] \psi \ . \eqno(4)$$
Although this seems to be trivial at the first sight, especially, when the
gauge field mass term and/or gauge-fixing
term for the symmetry (2) are absent, the consequences are not
$^{3, 16-18}$. The reason is that that quantization of a chiral fermion
results in a non-trivial interaction that breaks the classical gauge symmetry
(anomaly). It is also possible to redefine the fermion field via

$$\psi \rightarrow e^{f\left( \phi , \gamma^{5}\right)}\psi $$

$${\bar \psi} \rightarrow e^{ f^{\dag}\left( \phi , \gamma^{5}\right)}{\bar
\psi}\ . $$
Then  the fermion field is not invariant with respect to (5). In fact, it is
also possible
to choose the function $f$ so that the scalar field $\phi$ is absent from
the Lagrangian (4). But then one should worry about the Jacobian in the
fermionic sector.
We have chosen the simplest field redefinition so that everything is explicit!
The transformation (6) introduces the following additional
Abelian gauge symmetry to the theory $^{3,16-18}$:

$$\delta \phi \left( x \right) = {\bar \alpha }\left( x \right) \eqno(5a)$$

$$\delta A_{\mu} \left( x \right) = - \int d^{n} x d^{n} y \left(
\frac{\delta g^{\mu} \left( A, \phi \right)}{\delta A_{\nu}} \right)^{-1}
\left( x,y \right) \frac{\delta g_{nu} \left( A,\phi \right) }{\delta \phi}
{\bar \alpha} \left( z \right) = \partial _{\mu}
{\bar \alpha} \left( x \right) \eqno(5b) $$

$$ \delta \psi = {\bar \psi } = 0 \ , \eqno(5c)$$
where ${\bar \alpha}$ is an arbitrary real function. To quantize this model
we have to fix both gauge symmetries $^{20}$.

\ \ \ Now, we are prepared to analyse the problem of quantization of an
anomalous chiral gauge theory. The authors of Ref. 13 and 14
proposed to perform the non-local transformation

$$A_{\mu} \rightarrow A^{g}_{\mu} = A_{\mu} - \partial _{\mu} \frac{1}{\Box}
\partial _{\nu} A^{\nu}  \eqno(6)$$
in (1). The resulting theory

$$ {\it L} = - \frac{1}{4} F^{\mu \nu} F_{\mu \nu} + {\bar \psi}\left[
i \partial ^{\mu} \gamma _{\mu} - \frac{e}{2} \left( 1 + \gamma ^{5} \right)
A_{\mu}^{g} \gamma ^{\mu} \right] \psi \ . \eqno(7)$$
is then invariant with respect to

$$\delta A_{\mu}=\partial _{\mu} \alpha  \eqno(8a)$$

$$\delta \psi = \delta {\bar \psi} = \delta A_{\mu}^{g} = 0 \ . \eqno(8b)$$
This symmetry
is anomaly-free because the fermion field transforms in a trivial way $^{21}$.
It can be shown that such non-local theories are unitary and consistent
$^{13-15}$. Unfortunately,  these  conclusions usually concern the
additional  gauge symmetry
that has been introduced to the theory in question, but not the one
we started with. The discussed above model is still anomalous with
respect to the original $U(1)$ gauge symmetry. Such a Lagrangian might
yield a physically acceptable theory, but this is far from being a rule
$^{1}$. We should get rid of the anomalous symmetry. The simplest solution
is the following. Let us try to quantize the model given by Eq.4.
First, let us break the original gauge
symmetry (2) by the non-local gauge fixing condition

$$ \phi - \frac{1}{\Box} \partial _{\mu} A^{\mu} = 0 \ . \eqno(9)$$
The Lagrangian has the form (we omit the Faddeev-Popov ghost term)

$$ {\it L} = - \frac{1}{4} F^{\mu \nu} F_{\mu \nu} + {\bar \psi}\left[
i \partial ^{\mu} \gamma _{\mu} - \frac{e}{2} \left( 1 + \gamma ^{5} \right)
\left( A_{\mu} \gamma ^{\mu} - \gamma ^{\mu}
\partial _{\mu} \phi \right) \right] \psi + \rho \left(
 \phi - \frac{1}{\Box} \partial _{\mu} A^{\mu} \right) \ , \eqno(10)$$
where an auxiliary scalar field $\rho$ has been introduced to
exponentiate the functional Dirac $\delta$-function that
force the gauge condition (9). We can perform the path integral over the
scalar fields. This results in

$$ {\it L} = - \frac{1}{4} F^{\mu \nu} F_{\mu \nu} + {\bar \psi}\left[
i \partial ^{\mu} \gamma _{\mu} - \frac{e}{2} \left( 1 + \gamma ^{5} \right)
\left( A_{\mu} \gamma ^{\mu} - \partial _{\nu} \gamma^{\nu}
\left( \frac{1}{\Box} \partial _{\mu} A^{\mu}\right)
\right) \right] \psi \ . \eqno(11) $$
This is the Lagrangian given by (7) $^{13}$ with the $A^{g}$
field written explicitly! The additional gauge symmetry (8) is the
same as (5). Of course, other gauge conditions lead to different
representation of the model. This shows that the  proposal  put  forward  in
Refs 13 and 14 is to break the original symmetry (2) and to introduce a new one
that is anomaly-free (and in some sense trivial because it leaves fermions
invariant). In fact, it can be
shown that the transformation (6) chooses the covariant gauge

$$\partial _{\mu} A^{\mu} = 0\ . \eqno(12)$$
So we should not speak of a transformation but rather of a gauge
fixing condition. More sophisticated gauge conditions breaking (5)
would result in more complicated Lagrangians.

\ \ \ Jackiw and Rajaraman, in their seminal paper $^{1}$,
discovered that the two dimensional chiral Schwinger model yields a consistent
and unitary, although anomalous and not gauge-invariant, theory.
After this, several other consistent anomalous models have been put forward.
They have the following general form

$$ {\it L} = - \frac{1}{4} F^{\mu \nu} F_{\mu \nu} + {\bar \psi}\left[
i \partial ^{\mu} \gamma _{\mu} - \frac{e}{2} \left( 1 + \gamma ^{5} \right)
A_{\mu} \gamma ^{\mu} \right] \psi $$
$$
+ \frac{1}{2} B^{2} - B\partial _{\mu}
A^{\mu} + \partial _{\mu} {\bar c} \partial ^{\mu} c +
m^{2} K\left( \phi, A \right) +
\phi P \left( A \right)
\ , \eqno(13)$$
where B , c and P denote the auxiliary field that linearize the gauge
condition, the appropriate ghosts and the Pontryagin term $^{22}$,
respectively. Several forms of the K-term have been discussed in the
literature $^{3, 1-7, 10-13}$. In the 1+1 dimensional case, it is
possible to calculate the functional integral over the fermions $^{22}$
in (1). Then one can apply the transformation (3) $^{3}$. This leads
(after "reintroduction"of  fermions and addition of the gauge
fixing and ghost terms) to

$$ m^{2}=\frac{e^{2}}{4\pi}\left( a-1 \right) \ ,\eqno(14a)$$
where $a$ is the quantization (regularization) ambiguity parameter [1]
and

$$K\left( \phi , A\right)= \frac{1}{2} \partial _{\mu} \phi
\partial ^{\mu} \phi -  \partial _{\mu} \phi A^{\mu} \ . \eqno(14b)$$
This form correspond to a theory that posses the additional gauge
symmetry (5). This additional symmetry is the unexpected gauge
invariance discovered in Ref. 12 after adding the Wess-Zumino term to the
chiral electrodynamics Lagrangian.
This form of the K-term has been
recognized in Ref. 14 as the one corresponding to the model discussed
by Jackiw and Rajaraman $^{1}$.
This not so, because the
additional symmetry is absent in  their model $^{3}$. One has to break the
additional symmetry in order to get the Jackiw and Rajaraman model $^{3}$.
Faddeev and Shatashvili $^{10}$ have chosen
$K=0$. This corresponds to the $a=1$ case of Eq. (14a). Path integration over
the scalar field $\phi$ leads to the condition $P=0$ that ensure
the invariance with respect to (5).
The form proposed by
Rajeev $^{11}$

$$K\left( \phi , A\right)= \frac{1}{2} \left( \partial _{\mu} \phi
-A_{\mu} \right)
 \left( \partial ^{\mu} \phi- A^{\mu} \right)\ , \eqno(15)$$
has an additional term $\frac{1}{2} e^{2} A_{\mu}A^{\mu}$ that breaks
the the symmetry (8). It should be interpreted as a mass term for the
gauge boson (St\" uckelbeg formalism $^{23}$).
Finally, Thompson and Zhang proposed to take

$$K\left( \phi , A\right)= \partial _{\mu} \rho
\left( \partial ^{\mu} \phi- A^{\mu} \right)\ , \eqno(16)$$
where  $\rho$  is  an  auxiliary  scalar  field.  This  model   is
equivalent to
the ordinary chiral Schwinger model $^{1,3}$. This can be seen by integrating
over the scalar fields. Note that this differs from (4) $^{13}$
in that the additional symmetry is broken. This shows once
more that for any value of a the original gauge symmetry is lost
in the quantization process $^{1, 22}$. The above analysis shows that the
consistency of quantization of the discussed models has common roots
that have been discovered by Jackiw and Rajaraman $^{1}$
because the differences in the K-terms can be regarded as different gauge
fixing terms for the symmetry (5). Note that the discussed Lagrangians
can be obtained also by more sophisticated ways $^{4-7, 24}$. The important
fact is that the additional symmetry (5) reveals itself in every case,
although it might  not be obvious, e.g. in the field-antifield formalism it
is fixed in the due process $^{5, 23- 25}$.\\

\ \ \ The important question {\it can a field-enlarging transformation help
to construct a non-trivial anomaly-free theory?} arises. The answer may be
affirmative. It has been observed that a theory can posses
a BRST symmetry $^{26}$ that is not a symmetry of the Lagrangian but only of
the
functional integral. This means that  several symmetries, if "broken
correctly",
may result in a anomaly-free subsymmetry (cancellation of the anomalous
terms in the fermionic determinant). To shed more light on the problem,
let us consider the BRST symmetries that correspond to (2) and (5).
The general formula for a BRST current associated to the fields that appear
in (13) is $^{14, 20, 26}$

$${\it J}^{\mu}_{BRST}=F^{\mu \nu}\partial _{\nu} c - \phi \frac{\partial P}
{\partial \partial _{\mu} A_{\nu}}\partial _{\nu}c + e{\it J}^{\mu}_{L}c +
B\partial ^{\mu} c
- \frac{\partial K}{\partial \partial_{\mu} \phi} \delta_{BRST} \phi \ ,
\eqno(17) $$
where ${\it J}_{L}$ denotes the left fermion current. Its divergence is

$$ \partial {\it J}_{BRST} = e\partial {\it J}_{L}c - P\delta _{BRST} \phi
-\left( \frac{\partial K}{\partial \partial A_{\mu}}\partial _{\mu} c +
\frac{\partial K}{\partial \partial_{\mu} \phi}\partial _{\mu}
\delta _{BRST}\phi \right) \ , \eqno(18)$$
so that if K is gauge invariant and $\delta _{BRST}\phi = c$ ,
${\it J}_{BRST}^{\mu}$ is conserved ($ \partial {\it J}_{BRST}=0$) $^{13}$.
It is obvious that this condition is fulfilled by the
K-terms given by (15) and (16).
This conserved BRST charge corresponds to the diagonal part of (2)
and (5) ($\alpha = {\bar \alpha}$). The form of the K-term given by (14)
(and its special case K=0) defines a model that is gauge invariant with
respect to (5) and the appropriate BRST current is also conserved.
Unfortunately, the described above phenomenon seems to require
additional fields or/and non-local terms. It is also obscure if, and to
what extent, it can work in more than 1+1 dimensional spacetime.
The Batalin-Vilkovisky or the discussed here field-enlarging (St\"uckelberg
$^{23}$)
formalism should be helpful in analysing this problem. Especially, the role
of the additional symmetry should be explored.
This problem is under investigation. Recently, similar ideas has been
discussed in the context of the $W_{2}-gravity$ $^{7}$.\\

\ \ \ {\bf Acknowledgements}. The author would like to thank prof. R. K\"
ogerler and dr K. Ko\l odziej for stimulating and helpful
discussions, and the anonymous Referee for helpful
remarks and bringing attention to the references 4 - 7. This work
has been supported in part by the Alexander von Humboldt Foundation and the
Polish Committee for Scientific Research under the contract KBN-PB 2253/2/91.

\newpage
\subsection*{\ \ References}

\newcounter{bban}

\begin{list}
{[\arabic{bban}]}{\usecounter{bban}\setlength{\rightmargin}
{\leftmargin}}

\item R. Jackiw and R. Rajaraman, Phys. Rev. Lett. {\bf 54}, 1219 (1985).
\item P. Mitra, Phys. Lett. {\bf B284}, 23 (1992).
\item J. S\l adkowski, Phys. Lett. {\bf B296}, 361 (1992).
\item O. Babelon F. A. Schaposnik and C. M. Viallet, Phys. Lett. {\bf B177},
385 (1986).
\item J. Gomis and J. Paris, Nucl. Phys {\bf B395}, 288 (1993).
\item N. R. F. Braga and L. Montani, Phys. Lett. {\bf B264}, 125 (1991).
\item F. De Jonghe, R. Siebelink and W. Troost, Phys Lett {\bf B 306},
295 (1993).
\item M. Abdelhafiz, K. Ko\l odziej and M. Zra\l ek, Jagellonian Univ.
preprint, TPJU-2/86 ,1986 (unpublished).
\item J. S\l adkowski and M. Zra\l ek, Phys. Rev. {\bf D45}, 1701 (1992).
\item L. D. Faddeev and S. L. Shatashvili, Phys. Lett. {\bf B167}, 225 (1986).
\item S. G. Rajeev, MIT Report No CTP 1405 (1986).
\item K. Harada and I. Tsutsui, Phys. Lett. {\bf B183}, 311 (1987).
\item G. Thompson and R. Zhang, Mod. Phys. Lett. {\bf A2}, 869 (1987).
\item A. Della Serva, L. Maspieri and G Thompson, Phys. Rev. {\bf D37}, 2347
(1998).
\item G. L. Demarco, C. Fosco and R. C. Trinchero, Int. J. Mod.
Phys. {\bf A7}, 5459 (1992).
\item J. Alfaro and P. H. Damgaard, Ann. Phys. (NY) {\bf 202}, 398 (1990).
\item J. Alfaro and P. H. Damgaard, CERN-preprint CERN-TH6455/92 (1992).
\item A. Hosoya and K. Kikkawa, Nucl. Phys. {\bf B101}, 271 (1975).
\item J. Bell and R. Jackiw, Nuovo Cim. {\bf 60A}, 47 (1969);
S. Adler, Phys. Rev. {\bf 177}, 2426 (1969).
\item T. Kugo and I. Ojima, Progr. Theor. Phys. Suppl. {\bf 66}, 1 (1979).
\item K. Fujikawa, Phys. Rev. {\bf D21}, 2848 (1980).
\item R. Jackiw, in {\it Relativity, Groups and Topology II}, edited by B.
DeWit
and R. Stora (North-Holland, Amsterdam, 1984), p. 224.
\item J. S\l adkowski, submitted to Zeitschr. Phys. {\bf C}.
\item J. Alfaro and P. H. Damgaard, CERN-preprint CERN-TH6788/93 (1993);
to apear in Nucl. Phys. {\bf B}.
\item T. Fujiwara, Y. Igarashi and J. Kubo, Nucl. Phys. {\bf B341}, 695 (1990).
\item C. Becci, A. Rouet, and R. Stora, Phys. Lett. {\bf B52}, 344 (1974);
I. V. Tyutin, Lebedev Report No. FIAN 39 (1975).

\end{list}

\end{document}